# The Energetics of Particle Acceleration Using High Intensity Lasers

A.C. Melissinos

*Department of Physics and Astronomy*

*University of Rochester, Rochester, NY 14627*

October 28, 2004

**Abstract**

We point out that even the most intense laser beams available today can provide only a very small fraction ($10^{-7}$) of the beam energy required to reach the design luminosity for a future $e^+e^-$ linear collider. This fact seems to have been overlooked in the extensive literature on laser acceleration of charged particles.

Since the early days of laser operation it was proposed that the associated electric field could be used to accelerate electrons or ions. Since the field in a freely propagating laser beam is principally transverse, acceleration cannot occur in vacuum [1] but the field must be modified by imposing boundaries such as structures that guide the field or by the formation of a plasma [2].

The goal of producing energetic electrons and ions using laser beams has been achieved in the past decade [3, 4, 5] and exploits the availability of very short laser pulses (sub-ps duration) of reasonable intensity [6]. For instance in [3] the emission of electrons with energies extending up to 200 MeV are reported. Typically few nC of charge are emitted, when integrated over all energies. The laser intensity was $I \sim 10^{18}$ W/cm$^2$ and the acceleration gradient deduced from the data, $dE/dx \sim 100$ GeV/m.

The observed accelerating gradient is almost three orders of magnitude larger than that achieved in radio-frequency linear accelerators where $dE/dx \sim 25-100$ MeV/m. However the length of the acceleration region achieved so far is extremely short as compared to typical linear accelerators which are several km long.

We wish to point out in this note that in spite of the high gradient, laser acceleration is not suitable for research in elementary particle physics. This is because even if the short acceleration length is overcome, lasers cannot deliver the energy required to reach the luminosity necessary for investigating new phenomena.

**Future Linear Collider**

To advance our present understanding of the elementary particles ever higher collision energies are needed. In the hadronic sector, the Fermilab $\overline{p}-p$ collider is the world's highest energy machine at a cm energy $\sqrt{s} = 2$ TeV [7]. The large hadron collider, to become operational at CERN in 2007, will collide protons on protons at $\sqrt{s} = 14$ TeV [8]. The only future accelerator relevant for particle physics is the proposed linear collider with $e^+$ and $e^-$ beams colliding at $\sqrt{s} = 1$ TeV [9]. Because electrons and positrons are not composed of quarks, an $e^+e^-$ collider provides new and complementary information to that which will be obtained from the LHC.

The effectiveness of a collider is measured by its luminosity [8]

$$L = \frac{N_e^2}{4\pi\sigma_x\sigma_y} n_b f_{rep} D \approx \frac{N_e^2}{A} f_c \qquad (1)$$

In the approximate expression for $L$ in Eq.(1), $N_e$ is the number of particles in one bunch (assumed the same for both beams), $A$ is the effective area of the beam and $f_c$ the frequency of collisions. To investigate processes of interest, the luminosity must be in the range

$$L \sim 10^{34} \text{ cm}^{-2}\text{s}^{-1} \qquad (2)$$

and the future LC is designed to achieve this goal. Typical parameters are

$$N_e = 10^{10}$$

$$E_e = 500 \text{ GeV}$$

$$A = 3 \times 10^4 \text{ nm}^2 \quad [\sigma_x = 500 \text{ nm} \quad \sigma_y = 5 \text{ nm}]$$

$$f_c = 1.5 \times 10^4 \text{ s}^{-1} \quad [n_b = 3000 \quad f_{rep} = 5 \text{ Hz}]$$

(We have set the disruption parameter $D = 1$)

Using the above parameters we can obtain the energy stored in a single bunch

$$U = eN_e E_e = 800 \text{ J} \qquad (3)$$

where $e$ is the electron charge. This energy must be supplied by the laser. The most powerful short pulse laser in operation today is one of the arms of NIF. This has delivered 20 kJ in the IR over a $\sim$ 10 ns pulse length [10].

We may assume that future research will make it possible to couple a significant fraction of the laser energy to the electron beam in order to satisfy the requirement of Eq.(3). The difficulty arises when we consider the repetition rate. The NIF lasers are based on glass amplifiers and can operate at a rate of $10^{-3}$ per second at best. On the other hand for a future LC the required repetition rate is $10^4$ Hz. Thus from pure energetic considerations a laser accelerator misses the requirements of high energy physics by a factor of $10^7$. This conclusion is independent of the method used for coupling the laser energy to the electron beam [11].

One could argue that by decreasing the beam area one could recover some fraction of the factor of $10^7$. This is not realistic since many other effects prevent the use of colliding beams of smaller areas than given above.